\pdfoutput=1
\documentclass[11pt,sigplan,nonacm]{acmart}
\usepackage[utf8]{inputenc}
\usepackage{booktabs}
\usepackage{natbib}
\usepackage{doi}
\usepackage{href-ul}
\usepackage{multirow}
\usepackage{mdframed}
\usepackage{tabularx}
\usepackage{tcolorbox}
\usepackage{paralist}
\usepackage{pgfplots}
\usepackage[capitalize]{cleveref}

\title{Developers' Perception: Fixed Bugs Often Overlooked as Quality Contributions}

\author{Vitaly Alifanov}
\email{alifanovv5@gmail.com}
\orcid{0009-0007-5226-9887}
\affiliation{\institution{Innopolis University}\city{Innopolis}\country{Russia}}

\author{Kamil Almetov}
\email{almetov.kamil@gmail.com}
\orcid{0009-0005-8174-1613}
\affiliation{\institution{Innopolis University}\city{Innopolis}\country{Russia}}

\author{Ivan Kornienko}
\email{thedarkknight.vanerk@gmail.com}
\orcid{0009-0000-8013-3839}
\affiliation{\institution{Innopolis University}\city{Innopolis}\country{Russia}}

\author{Arsen Mutalapov}
\email{arsenmutalapov205@gmail.com}
\orcid{0009-0000-4578-8944}
\affiliation{\institution{Innopolis University}\city{Innopolis}\country{Russia}}

\author{Yegor Bugayenko}
\orcid{0000-0001-6370-0678}
\email{yegor256@gmail.com}
\affiliation{\institution{Huawei}\city{Moscow}\country{Russia}}

\pgfplotsset{compat=1.18}
\begin{document}

\begin{abstract}
High-quality software products rely on both well-written source code and timely detection and thorough reporting of bugs. However, some programmers view bug reports as negative assessments of their work, leading them to withhold reporting bugs, thereby detrimentally impacting projects. Through a survey of 102 programmers, we discovered that only a third of them perceive the quantity of bugs found and rectified in a repository as indicative of higher quality. This finding substantiates the notion that programmers often misinterpret the significance of testing and bug reporting.
\end{abstract}

\maketitle

\section{Introduction}

Programmers contribute to the creation of a software product by writing code, while testers contribute by finding and reporting bugs~\citep{bugayenko2018discovering}. The importance of testing is hard to underestimate, especially when the ultimate quality of service and reliability of a product are important~\citep{black2016pragmatic,beizer1995black}. The primary output of testing is bug reports, while their quantity and quality may be used as indicators of the effectiveness of testing~\citep{gelperin1988growth,beizer2003software,bugayenko2018codeahead}.

However, not every programmer understands the importance of testing and the value that a large number of high-quality reports may bring to a project~\citep{pettichord2013lessons}. Some programmers may perceive bug reports as a negative opinion about their work~\citep{bugayenko2018codeahead,bugayenko2015blog0618}. Because of this, they may refrain from submitting them, trying not to hurt the feelings of other programmers~\citep{zhang2014sources}. Some programmers may believe that the absence of bug reports in their projects is a sign of high quality of coding~\citep{hutcheson2003software}. Such an attitude of programmers may hurt software projects~\citep{cohen2004managing}. Even though some authors raised this concern, there were no empirical studies conducted so far that would help understand how many programmers share this attitude.

In order to shed light on this problem, we interviewed 102 programmers with a two-phased survey. In the first phase, we asked them to choose a subjectively ``better'' software product out of five available, showing them ten metrics for each project: the ``Closed Bug Reports'' was one of the metrics. We also asked them which metric they pay attention to. The target metric appeared to not even be in the top three. In the second phase, we showed them a sample low-quality bug report and asked what they would do with it, giving them six options to choose from: three options were in favor of fixing the report, while another three were in favor of not reporting it at all. More than 60\% of answers were in favor of improving and reporting, while 40\% suggested not reporting it at all.

The results demonstrate that programmers are not afraid of reporting bugs but don't consider a larger number of bug reports in a project as an indicator of its higher quality. While the first finding is positive, the second one may be interpreted as a confirmation of the belief that programmers don't understand the value of testing.

The rest of the paper is organized as following.
\Cref{sec:related} shows a few previous studies related to the problem we reseach.
\Cref{sec:method} explains our method of study.
\Cref{sec:results} presents the results we obtained.
\Cref{sec:discussion} discusses our findings and highlights limitations of our method.

\section{Related Work}
\label{sec:related}

Software quality is being researched thoroughly, and the research-related studies are divided into several parts such as testing, debugging, and developers' opinions.

\citet{Testing} studied whether developers, consultants, testers, and managers want to test their software. The results show that developers think of testing as mundane and tend to prioritize other tasks. Thus, one's opinion on testing does not only rely on software quality but rather on personal motivation to improve the software.

\citet{APR} considered 386 software developers' opinions on automatic program repair tools (APR). One of the results was that programmers tend to use APR in testing to find bugs faster. Still, developers want to have either choice or control over the bug-fixing process because accuracy is a significant concern for most developers. Despite that, analysis of APR does not cover the perceptions toward software quality, but rather, a portion of it. Thus, a comprehensive study is still needed.

\citet{BugChanges} asked 41 participants to answer three research questions regarding developers' viewpoints about bug-introducing changes, reasons to agree with an assumption, and which assumptions are consensus/dissensus among viewpoints. The results about bug-introducing changes drew five opinions, and the most popular of them was ``no matter the software history, only tests, and developers’ experience are important'' with the belief that experienced developers introduce fewer bugs. However, the survey did not include the study of views about bugs, rather only about the consequences of bugs and low software quality.

To conclude, the research community produces papers and surveys on software quality, testing, and debugging frequently. However, to the best of our knowledge, there was no research on the perception of programmers on the bug reporting process and its relation to the software quality.

\section{Method}
\label{sec:method}

The main objective of the research was to determine how programmers perceive bug reports. It led us to the following research questions:
\begin{description}
  \item[RQ1] Do programmers consider software quality being positively correlated with the number of bugs reported?
  \item[RQ2] Do programmers feel comfortable when they report bugs?
\end{description}
We created a two-phased survey, where each phased helped us answer one RQ.

In the first phase of the survey, we asked programmers to choose one out of six projects. We asked them to select the project that, according to their subjective understanding, has the highest quality. For each project, we showed them ten metrics, as shown in \cref{tab:frameworks}. We did show them the names of the projects but did not provide links to GitHub repositories. Thus, they were unable to study the source code of the projects and had to rely solely on the provided metrics while making their decisions.

We also asked programmers to explain their choice by identifying the most important metric out of ten available. The answers were intended to reveal the importance of the number of closed bug reports (one of the metrics on the list), although this was not explicitly announced to the respondents. Those programmers who chose this metric would be considered as those who perceive the number of reported and fixed bugs as being positively correlated with software quality.

\begin{table*}
    \begin{tabular}{l *{5}{r}}
    \toprule
    \multicolumn{1}{c}{Metric} & \multicolumn{1}{c}{xgen} & \multicolumn{1}{c}{SWXMLHash}
    & \multicolumn{1}{c}{Fuzi} & \multicolumn{1}{c}{quick-xml} & \multicolumn{1}{c}{libexpat} \\
    \midrule
    Closed bug reports & 2 & 29 & 1 & 64 & 79 \\
    GitHub Stars & 276 & 1{,}345 & 1{,}049 & 982 & 918 \\
    Lines of code & 3{,}777 & 4{,}407 & 7{,}027 & 17{,}837 & 652{,}333 \\
    Release date (days ago)& 1{,}040 & 553 & 1{,}062 & 2{,}840 & 382 \\
    Contributors & 11 & 38 & 21 & 86 & 83 \\
    Closed issues & 18 & 140 & 57 & 264 & 286 \\
    \multicolumn{1}{l}{Primary programming language} & \multicolumn{1}{l}{Go} &
    \multicolumn{1}{l}{Swift} & \multicolumn{1}{l}{Swift} & \multicolumn{1}{l}{Rust} & \multicolumn{1}{l}{C} \\
    Last commit (days ago) & 126 & 179 & 176 & 2 & 10 \\
    GitHub Forks & 66 & 196 & 149 & 205 & 422 \\
    Closed pull requests & 22 & 114 & 45 & 333 & 466 \\
    \bottomrule
    \end{tabular}
\caption{This is a comparison table that contains XML parsing frameworks and metrics for them taken from GitHub. Respondents should put these libraries in the order from the most preferable to the least. Then, they should explain their choice and provide key metrics developers were based on.}
\label{tab:frameworks}
\end{table*}

At the second phase of the survey, we asked programmers to study a sample bug report, which intentionally contained a number of quality issues, as demonstrated in \cref{fig:fig1}. We gave programmers six options, which were implicitly split into two groups:
\begin{inparaenum}[1)]
    \item real problems with the bug report,
    and
    \item ``emotional`` barriers that a programmer could and should ignore.
\end{inparaenum}
Choosing answers from the second group means that they feel uncomfortable while reporting bugs. The options sound like this (the group names were not visible to programmers):
\begin{enumerate}[Op1:]
    \item (G1) Better explain how to reproduce it
    \item (G2) Don't report, since it's a primitive bug
    \item (G1) Get rid of the redundant 3--minute video
    \item (G2) Instead, post to project chat, for a faster fix
    \item (G1) Specify its priority and severity
    \item (G2) RTFM, maybe you do something wrong
\end{enumerate}

\begin{figure}
\begin{mdframed}\ttfamily
Error during login:\\
Message with some errors is displayed when attempting to log in. No authentication token is obtained. I tried to log in with a valid username and password, but no auth token was received. I used logs to check the token\\[.5em]
To reproduce try to log in with:\\
Login="login123"\\
Password="password123"\\[.5em]
System information:\\
Browser=Google Chrome\\[.5em]
Attachments:\\
3-minute video with authorization process\\[.5em]
Additional information:\\
This issue started occurring after the latest system update. I have tried logging in on multiple devices and browsers, and the issue persists.
\end{mdframed}
\caption{This is the bug report about some authentication problems. Respondents were asked to choose a few options out of six: what would they do in order to improve this report?}
\label{fig:fig1}
\end{figure}

Our intention was to completely conceal our research objective from the respondents. The question in the first phase included nine metrics besides bug reporting. The question in the second phase didn't directly inquire about feelings---it asked about the most harmful aspects of the particular bug report. We believe that concealing the research questions from the respondents enhanced the objectivity of the results. Moreover, the questions are situational, as they engage and interest the respondent more effectively in answering.

In compiling the questions, we were guided by the criteria in the related book~\citep{QuestionnaireBook}. In particular, we were guided by the brevity and simplicity of the questions to take the least time for the interviewee to reach as large an audience as possible.

\section{Results}
\label{sec:results}

Using our existing professional connections and social networks, we invited programmers to participate in the survey, through Google Forms. 102 people agreed to participate, knowing that the results of this study will be published\footnote{The entire CSV dataset with all answers provided is available in a public GitHub repository: \url{https://github.com/system205/QuestionnaireData}}. 80 of them filled up the form in English, while 22 in Russian.

\Cref{tab:positions} demonstrates the distribution of positions of respondents in their companies, as they reported in the survey. The majority (55\%) of them identified themselves as engineers (``Developer,'' ``Engineer,'' and ``Senior''), which is suitable for the goals of our research.
\begin{table}
\begin{tabular}{lrr}
\toprule
Position & Responses &  \\
\midrule
Developer & 15 & 37\%\\
Lead/Manager/Head/Chief & 6 & 15\%\\
Senior & 4 & 10\%\\
Engineer & 3 & 8\%\\
Intern & 2 & 5\%\\
Student & 2 & 5\%\\
Junior & 1 & 3\%\\
Others& 7 & 17\%\\
\bottomrule
\end{tabular}
\caption{A summary of work positions of respondents, as they reported in the survey, where ``Others'' include responses like ``Freelance,'' ``PM,'' ``Lab,'' ``Unknown'' and no responses at all. The total count of responses is not equal to the total number of respondents because not everybody answered to this question in the survey.}
\label{tab:positions}
\end{table}

\Cref{tab:experience} demonstrates how experienced are the respondents, according to their own information provided while conducting the survey. Almost the majority of respondents (46\%) are rather experienced software engineers (three or more years), which is suitable for the goals of our research.
\begin{table}
\begin{tabular}{lrr}
\toprule
Years of experience & Responses & \\
\midrule
0 \(\leq\) n < 1 & 13 & 27\%\\
1 \(\leq\) n < 3 & 9 & 19\%\\
3 \(\leq\) n < 5 & 6 & 13\%\\
5 \(\leq\) n < 10 & 10 & 20\%\\
n \(\geq\) 10 & 6 & 13\%\\
Others & 4 & 8\%\\
\bottomrule
\end{tabular}
\caption{A summary of years of experience, as they were claimed by the respondents. The total count of responses is not equal to the total number of respondents because not everybody answered to this question in the survey.}
\label{tab:experience}
\end{table}

\Cref{tab:reported} demonstrates the distribution of the amount of bugs respondents report per year, according to their own evaluation. The majority of them (73\%) reported less than 15 bugs, which is one bug per month. This number looks reasonably low and may confirm the intuition that programmers tend to avoid formal reporting of bugs.
\begin{table}
\begin{tabular}{lrr}
\toprule
Number of bugs & Responses & \\
\midrule
0 & 6 & 17\%\\
0 < n < 5 & 8 & 23\%\\
5 \(\leq\) n < 15 & 12 & 33\%\\
15 \(\leq\) n < 40 & 3 & 8\%\\
40 \(\leq\) n < 100 & 4 & 11\%\\
Others & 3 & 8\%\\
\bottomrule
\end{tabular}
\caption{A summary of how many bugs respondents reported during the last calendar year. The total count of responses is not equal to the total number of respondents because not everybody answered to this question in the survey.}
\label{tab:reported}
\end{table}

\subsection{RQ1: Do programmers consider software quality being positively correlated with the number of bugs reported?}

Survey respondents were asked to place different XML parsing libraries from most preferable to the least. The resulting distribution is shown in \cref{tab:tab2}. The results demonstrate that libraries with more closed bug reports and closed issues (``libexpat'' and ``quick-xml'') are preferable. However, when we asked respondents to answer ``What are the metrics you were based on?,'' we revealed that the main metrics are the ``number of stars,'' ``primary programming language,'' and ``number of days from the last commit'' (``quick-xml'' and ``libexpat'' have the smallest numbers: two and ten days respectively).

The distribution of chosen metrics is shown in \cref{tab:tab3}, indicating how many of them were reported by respondent as the first or primary choice. Even though 34 respondents considered bug reports and issues when choosing the library with the higher quality, these metrics are not determinant taking only the 4th and the 7th places among primary factors.

\begin{table}
\begin{tabular}{l *{5}{r}}
\toprule
\multicolumn{1}{c}{Framework} &  \multicolumn{1}{c}{Top1} &  \multicolumn{1}{c}{Top2} &  \multicolumn{1}{c}{Top3} &  \multicolumn{1}{c}{Top4} &  \multicolumn{1}{c}{Top5} \\
\midrule
libexpat & 28 & 28 & 13 & 9 & 24 \\
quick-xml & 36 & 22 & 20 & 17 & 7 \\
SWXMLHash & 25 & 26 & 26 & 15 & 10 \\
xgen & 8 & 10 & 17 & 20 & 47 \\
Fuzi & 5 & 16 & 26 & 41 & 14 \\
\bottomrule
\end{tabular}
\caption{This table demonstrates metrics and their priorities. Top1 indicates the highest priority, and Top5 means the least priority.}
\label{tab:tab2}
\end{table}

\begin{table}
\begin{tabular}{lrr}
\toprule
 \multicolumn{1}{c}{Metric} &  \multicolumn{1}{c}{Top priority} &  \multicolumn{1}{c}{Rest} \\
\midrule
Stars & 28 & 26 \\
Primary language & 21 & 21 \\
Last commit (days ago) & 11 & 34 \\
Release date (days ago) & 8 & 23 \\
Closed bug reports & 8 & 26 \\
Lines of code & 6 & 26 \\
Contributors & 5 & 29 \\
Closed issues & 5 & 29 \\
Forks & 4 & 19 \\
Closed pull requests & 1 & 10 \\
\bottomrule
\end{tabular}
\caption{This table produces how a specific metric is chosen as the highest priority and the rest kinds of priorities denoted as ``Rest.''}
\label{tab:tab3}
\end{table}

\begin{table}
\begin{tabular}{lrr}
\toprule
Option & Votes & \\
\midrule
Op1: Explain how to reproduce it & 60 & 25\%\\
Op2: Don’t report a primitive bug & 4 & 2\%\\
Op3: Remove the 3–minute video & 42 & 18\%\\
Op4: Post to project chat & 20 & 8\%\\
Op5: Specify priority and severity & 66 & 27\%\\
Op6: RTFM, maybe you are wrong & 49 & 20\%\\
\bottomrule
\end{tabular}
\caption{A summary of votes our respondents provided for each option. The total count of votes is larger than the total number of respondents because every respondent was allowed to chose more than one option.}
\label{tab:tab4}
\end{table}

As an answer to RQ1, we conclude that programmers consider software quality to be positively correlated with the number of bugs reported but the correlation between quality and number of bugs is not strong. Exactly 33\% of respondents can answer ``yes'' to RQ1. The respondents' answers demonstrate that developers pay more attention to other metrics such as  ``programming language'' and ``number of days since the last commit.''

\subsection{RQ2: Do programmers feel comfortable when they report bugs?}

We showed a bug report related to an authentication problem (\cref{fig:fig1}) and provided six possible options to choose from. Three of them were focused on specific improvements and further reporting, and another three were related to the avoidance of reporting.

\Cref{tab:tab4} shows the distribution of chosen options on the request ``What would you do to improve the bug report?'' In total, we got 2.5 times more votes about technical improvements (167) rather than mental avoidance (68). Thus, the majority of voters (60\%) are okay with the reporting of this bug. In addition, 40\% of respondents chose only technical options.

Summarizing these results, we can state that programmers are not afraid of bug reporting. However, about half of them feel uncertain and prefer to read a manual before submitting a report.

\section{Discussion}
\label{sec:discussion}

\textbf{Why do most programmers choose GitHub stars as a metric of quality?}
Indeed, even though we emphasized quality, some developers referred to the term ``popularity.'' The choice of popularity and GitHub stars might be because developers shift the responsibility of choice to others, even though we asked them to make their selection based on quality.

\textbf{Is 102 respondents a large enough study group?}
While other researchers have asked about respondents' opinions and viewpoints, our method included situational questions which elicited more truthful answers. Due to this methodology, respondents treat a question as a real-life situation, which is a key strength of our study. However, there are limitations. One of them is the number of participants. 102 people represent a relatively small sample to describe the common thinking of software developers. In future empirical studies, we plan to interview a larger group of engineers.

\textbf{How reliable is the statistics collected through the opinions of the bug report?}
Indeed, the list of options we provided to the respondents is rather limited. People may rely on different behaviors not listed in the specified options, and we do not provide an opportunity for them to write their own opinion. We also found that about half of the programmers did not feel any ambiguity when dealing with external bug reports.

\section{Conclusion}
\label{sec:conclusion}

We conducted a survey and studied the attitude of 102 software professionals towards bug reporting and its impact on software quality. Along with our initial hypothesis, only a third of all respondents suggested considering the amount of reported and fixed bug reports in a repository as an indicator of its quality. However, contrary to our initial hypothesis, the survey demonstrated that programmers do not feel bad about reporting bugs. Thus, we conclude that most programmers do not fully understand the importance and the value that testing and bug reporting provide to their projects.

{\raggedright\bibliographystyle{ACM-Reference-Format}
\bibliography{main}}


\begin{thebibliography}{15}


\ifx \showCODEN    \undefined \def \showCODEN     #1{\unskip}     \fi
\ifx \showDOI      \undefined \def \showDOI       #1{#1}\fi
\ifx \showISBNx    \undefined \def \showISBNx     #1{\unskip}     \fi
\ifx \showISBNxiii \undefined \def \showISBNxiii  #1{\unskip}     \fi
\ifx \showISSN     \undefined \def \showISSN      #1{\unskip}     \fi
\ifx \showLCCN     \undefined \def \showLCCN      #1{\unskip}     \fi
\ifx \shownote     \undefined \def \shownote      #1{#1}          \fi
\ifx \showarticletitle \undefined \def \showarticletitle #1{#1}   \fi
\ifx \showURL      \undefined \def \showURL       {\relax}        \fi
\providecommand\bibfield[2]{#2}
\providecommand\bibinfo[2]{#2}
\providecommand\natexlab[1]{#1}
\providecommand\showeprint[2][]{arXiv:#2}

\bibitem[Beizer(1995)]%
        {beizer1995black}
\bibfield{author}{\bibinfo{person}{Boris Beizer}.}
  \bibinfo{year}{1995}\natexlab{}.
\newblock \bibinfo{booktitle}{\emph{{Black-Box Testing: Techniques for
  Functional Testing of Software and Systems}}}.
\newblock \bibinfo{publisher}{John Wiley \& Sons}.
\newblock


\bibitem[Beizer(2003)]%
        {beizer2003software}
\bibfield{author}{\bibinfo{person}{Boris Beizer}.}
  \bibinfo{year}{2003}\natexlab{}.
\newblock \bibinfo{booktitle}{\emph{{Software Testing Techniques}}}.
\newblock \bibinfo{publisher}{Dreamtech Press}.
\newblock


\bibitem[Black(2016)]%
        {black2016pragmatic}
\bibfield{author}{\bibinfo{person}{Rex Black}.}
  \bibinfo{year}{2016}\natexlab{}.
\newblock \bibinfo{booktitle}{\emph{{Pragmatic Software Testing: Becoming an
  Effective and Efficient Test Professional}}}.
\newblock \bibinfo{publisher}{John Wiley \& Sons}.
\newblock


\bibitem[Bugayenko(2015)]%
        {bugayenko2015blog0618}
\bibfield{author}{\bibinfo{person}{Yegor Bugayenko}.}
  \bibinfo{year}{2015}\natexlab{}.
\newblock \bibinfo{title}{{Good Programmers Write Bug-Free Code, Don't They?}}
\newblock \bibinfo{howpublished}{\url{https://www.yegor256.com/150618.html}}.
\newblock
\newblock
\shownote{[Online; accessed 05-03-2024]}.


\bibitem[Bugayenko(2018a)]%
        {bugayenko2018codeahead}
\bibfield{author}{\bibinfo{person}{Yegor Bugayenko}.}
  \bibinfo{year}{2018}\natexlab{a}.
\newblock \bibinfo{booktitle}{\emph{{Code Ahead}}}.
\newblock \bibinfo{publisher}{Amazon}.
\newblock


\bibitem[Bugayenko(2018b)]%
        {bugayenko2018discovering}
\bibfield{author}{\bibinfo{person}{Yegor Bugayenko}.}
  \bibinfo{year}{2018}\natexlab{b}.
\newblock \showarticletitle{{Discovering Bugs, or Ensuring Success?}}
\newblock \bibinfo{journal}{\emph{{Communications of the ACM}}}
  \bibinfo{volume}{61}, \bibinfo{number}{9} (\bibinfo{year}{2018}),
  \bibinfo{pages}{12--13}.
\newblock
\urldef\tempurl%
\url{https://doi.org/10.1145/3237196}
\showDOI{\tempurl}


\bibitem[Cohen et~al\mbox{.}(2004)]%
        {cohen2004managing}
\bibfield{author}{\bibinfo{person}{Cynthia~F. Cohen},
  \bibinfo{person}{Stanley~J. Birkin}, \bibinfo{person}{Monica~J. Garfield},
  {and} \bibinfo{person}{Harold~W. Webb}.} \bibinfo{year}{2004}\natexlab{}.
\newblock \showarticletitle{{Managing Conflict in Software Testing}}.
\newblock \bibinfo{journal}{\emph{{Communications of the ACM}}}
  \bibinfo{volume}{47}, \bibinfo{number}{1} (\bibinfo{year}{2004}),
  \bibinfo{pages}{76--81}.
\newblock
\urldef\tempurl%
\url{https://doi.org/10.1145/962081.962083}
\showDOI{\tempurl}


\bibitem[Gelperin and Hetzel(1988)]%
        {gelperin1988growth}
\bibfield{author}{\bibinfo{person}{David Gelperin} {and} \bibinfo{person}{Bill
  Hetzel}.} \bibinfo{year}{1988}\natexlab{}.
\newblock \showarticletitle{{The Growth of Software Testing}}.
\newblock \bibinfo{journal}{\emph{{Communications of the ACM}}}
  \bibinfo{volume}{31}, \bibinfo{number}{6} (\bibinfo{year}{1988}),
  \bibinfo{pages}{687--695}.
\newblock
\urldef\tempurl%
\url{https://doi.org/10.1145/62959.62965}
\showDOI{\tempurl}


\bibitem[Gillham(2008)]%
        {QuestionnaireBook}
\bibfield{author}{\bibinfo{person}{Gillham}.} \bibinfo{year}{2008}\natexlab{}.
\newblock \bibinfo{booktitle}{\emph{{Developing a Questionnaire (2nd Ed.)}}}.
\newblock \bibinfo{publisher}{Continuum}.
\newblock


\bibitem[Hutcheson(2003)]%
        {hutcheson2003software}
\bibfield{author}{\bibinfo{person}{Marnie~L. Hutcheson}.}
  \bibinfo{year}{2003}\natexlab{}.
\newblock \bibinfo{booktitle}{\emph{{Software Testing Fundamentals: Methods and
  Metrics}}}.
\newblock \bibinfo{publisher}{John Wiley \& Sons}.
\newblock
\urldef\tempurl%
\url{https://doi.org/10.5555/862109}
\showDOI{\tempurl}


\bibitem[Pettichord et~al\mbox{.}(2013)]%
        {pettichord2013lessons}
\bibfield{author}{\bibinfo{person}{Bret Pettichord}, \bibinfo{person}{James
  Bach}, {and} \bibinfo{person}{Cem Kaner}.} \bibinfo{year}{2013}\natexlab{}.
\newblock \bibinfo{booktitle}{\emph{{Lessons Learned in Software Testing: A
  Context-Driven Approach}}}.
\newblock \bibinfo{publisher}{Wiley}.
\newblock


\bibitem[Souza et~al\mbox{.}(2022)]%
        {BugChanges}
\bibfield{author}{\bibinfo{person}{Jairo Souza}, \bibinfo{person}{Rodrigo
  Lima}, \bibinfo{person}{Baldoino Fonseca}, \bibinfo{person}{Bruno Cartaxo},
  \bibinfo{person}{M\'{a}rcio Ribeiro}, \bibinfo{person}{Gustavo Pinto},
  \bibinfo{person}{Rohit Gheyi}, {and} \bibinfo{person}{Alessandro Garcia}.}
  \bibinfo{year}{2022}\natexlab{}.
\newblock \showarticletitle{{Developers' Viewpoints to Avoid Bug-Introducing
  Changes}}.
\newblock \bibinfo{journal}{\emph{{Information and Software Technology}}}
  \bibinfo{volume}{143}, \bibinfo{number}{1} (\bibinfo{year}{2022}).
\newblock
\urldef\tempurl%
\url{https://doi.org/10.1016/j.infsof.2021.106766}
\showDOI{\tempurl}


\bibitem[Straubinger and Fraser(2023)]%
        {Testing}
\bibfield{author}{\bibinfo{person}{Philipp Straubinger} {and}
  \bibinfo{person}{Gordon Fraser}.} \bibinfo{year}{2023}\natexlab{}.
\newblock \showarticletitle{{A Survey on What Developers Think About Testing}}.
  In \bibinfo{booktitle}{\emph{{Proceedings of the 34th International Symposium
  on Software Reliability Engineering (ISSRE)}}}. \bibinfo{pages}{80--90}.
\newblock
\urldef\tempurl%
\url{https://doi.org/10.1109/ISSRE59848.2023.00075}
\showDOI{\tempurl}


\bibitem[Winter et~al\mbox{.}(2023)]%
        {APR}
\bibfield{author}{\bibinfo{person}{Emily Winter}, \bibinfo{person}{David
  Bowes}, \bibinfo{person}{Steve Counsell}, \bibinfo{person}{Tracy Hall},
  \bibinfo{person}{Saemundur Haraldsson}, \bibinfo{person}{Vesna Nowack}, {and}
  \bibinfo{person}{John Woodward}.} \bibinfo{year}{2023}\natexlab{}.
\newblock \showarticletitle{{How Do Developers Really Feel About Bug Fixing?
  Directions for Automatic Program Repair}}.
\newblock \bibinfo{journal}{\emph{{IEEE Transactions on Software Engineering}}}
  \bibinfo{volume}{49}, \bibinfo{number}{4} (\bibinfo{year}{2023}),
  \bibinfo{pages}{1823--1841}.
\newblock
\urldef\tempurl%
\url{https://doi.org/10.1109/TSE.2022.3194188}
\showDOI{\tempurl}


\bibitem[Zhang et~al\mbox{.}(2014)]%
        {zhang2014sources}
\bibfield{author}{\bibinfo{person}{Xihui Zhang}, \bibinfo{person}{Thomas~F.
  Stafford}, \bibinfo{person}{Jasbir~S. Dhaliwal}, \bibinfo{person}{Mark~L.
  Gillenson}, {and} \bibinfo{person}{Gertrude Moeller}.}
  \bibinfo{year}{2014}\natexlab{}.
\newblock \showarticletitle{{Sources of Conflict Between Developers and Testers
  in Software Development}}.
\newblock \bibinfo{journal}{\emph{{Information \& Management}}}
  \bibinfo{volume}{51}, \bibinfo{number}{1} (\bibinfo{year}{2014}),
  \bibinfo{pages}{13--26}.
\newblock
\urldef\tempurl%
\url{https://doi.org/10.1016/j.im.2013.09.006}
\showDOI{\tempurl}


\end{thebibliography}

\end{document}